\documentclass[aps,pra,twocolumn,amsmath,amssymb,showpacs,superscriptaddress]{revtex4-1}
\usepackage{amsmath,amsfonts,amssymb}
\usepackage{physics}
\usepackage{graphicx}
\usepackage{setspace}
\usepackage{tocloft}
\usepackage[dvipsnames]{xcolor}

\cftpagenumbersoff{figure}
\cftpagenumbersoff{table}

\begin{document} 
\title{Robust structured light in atmospheric turbulence}

\author{Asher Klug}
\affiliation{School of Physics, University of the Witwatersrand, Private Bag 3, Wits 2050, South Africa}

\author{Cade Peters}
\affiliation{School of Physics, University of the Witwatersrand, Private Bag 3, Wits 2050, South Africa}

\author{Andrew Forbes}
\email[email:]{ andrew.forbes@wits.ac.za}
\affiliation{School of Physics, University of the Witwatersrand, Private Bag 3, Wits 2050, South Africa}
\email[Corresponding author: ]{andrew.forbes@wits.ac.za}

\date{\today}

\begin{abstract}
\noindent Structured light is routinely used in free space optical communication channels, both classical and quantum, where information is encoded in the spatial structure of the mode for increased bandwidth. Unlike polarisation, the spatial structure of light is perturbed through such channels by atmospheric turbulence, and consequently, much attention has focused on whether one mode type is more robust than another, but with seemingly inconclusive and contradictory results.  Both real-world and experimentally simulated turbulence conditions have revealed that free-space structured light modes are perturbed in some manner by turbulence, resulting in both amplitude and phase distortions. Here, we present complex forms of structured light which are invariant under propagation through the atmosphere: the true eigenmodes of atmospheric turbulence. We provide a theoretical procedure for obtaining these eigenmodes and confirm their invariance both numerically and experimentally.  Although we have demonstrated the approach on atmospheric turbulence, its generality allows it to be extended to other channels too, such as underwater and in optical fibre. 
\end{abstract}

\maketitle

% Include a list of up to six keywords after the abstract
%\keywords{structured light, atmospheric turbulence, invariant modes, free space optical communication}

% Include email contact information for corresponding author
%{\noindent \footnotesize\textbf{*}Corresponding author,  \linkable{andrew.forbes@wits.ac.za} }

\section{Introduction}\label{sec:intro}
Free-space transmission of electromagnetic waves is crucial in many diverse applications, including sensing, detection and ranging, defence and communication, and extends over distances from the long (Earth monitoring) to the short (WiFi and LiFi).  Lately there has been a resurgence of interest in free-space optical links \cite{trichili2019communicating,trichili2020roadmap}, driven in part by the need for increased communication bandwidths \cite{Richardson1,richardson2013space}, with the potential to bridge the digital divide in a manner that is license-free \cite{lavery2018tackling}. Here the spatial modes of light have come to the fore, for so-called space division multiplexing \cite{li2014space} and mode division multiplexing \cite{Berdague1982A}, where the spatial structure of light is used as an encoding degree of freedom.  This in turn has fuelled interest in structured light \cite{forbes2021structured,forbes2020structured}, where light is tailored in all its degrees of freedom, including amplitude, phase and polarisation, enabled by a modern structured light toolkit \cite{shen2021rays}.

A commonly used form of structured light is that of beams carrying orbital angular momentum (OAM), where the phase spirals around the path of propagation azimuthally \cite{padgett2017orbital}. These modes provide a (theoretically) infinite and easily realised alphabet for encoding information \cite{Gibson2004,wang2012terabit} and have been used extensively in optical communication (see Refs.~\cite{Willner2015,willner2021orbital} for good reviews). Vectorial combinations of such beams create inhomogeneous polarisation structures \cite{zhan2009cylindrical,rosales2018review,otte2018polarization} and too have found applications in free-space links \cite{nape2021revealing,Milione2015d,Sit2017}.  Although these structured light fields hold tremendous potential for free-space optical communication, they are distorted by atmospheric turbulence as a phase perturbation in the near-field and an amplitude, phase and polarisation perturbation in the far-field \cite{nape2021revealing}.  This modal scattering induced cross-talk decreases the information capacity of classical atmospheric transmission channels \cite{JaimeA.Anguita2008,ren2016experimental,krenn2014communication,zhao2016experimental,rodenburg2012a,krenn2016twisted,zhang2020mode,malik2012influence,chen2016changes,tyler2009influence} while reducing the degree of entanglement in quantum links \cite{Paterson2005,jha2010c,Tyler2009,oamturb,gopaul2007effect,zhang2016experimentally,pors2011transport,goyal2016effect,leonhard2015universal,sorelli2019entanglement}. Mitigating this remains an open challenge that is intensely studied.   

Arguments have been put forward for one mode family being more robust than another, with studies covering Bessel-Gaussian~\cite{mphuthi2018bessel,mphuthi2019free,lukin2014mean,bao2009propagation,zhu2008propagation,nelson2014propagation,ahmed2016mode,cheng2016channel,doster2016laguerre,watkins2020experimental,vetter2019realization,yuan2017beam}, Hermite-Gaussian~\cite{cox2019hglg,ndagano2017c,Restuccia2016,ndagano2017comparing}, Laguerre-Gaussian~\cite{Trichili2016,zhao2015capacity,zhou2019using,xie2016experimental,li2017power} and Ince-Gaussian~\cite{krenn2019turbulence} beams, with mixed and contradictory results. In the context of OAM, since the atmosphere itself can be thought of giving or taking OAM from the beam, it has been shown theoretically and experimentally that atmospheric turbulence distortions are independent of the original OAM mode \cite{klug2021orbital}, all susceptible to the deleterious effects of atmospheric turbulence, and indeed OAM has been suggested as not the ideal modal carrier through turbulence \cite{miller2017better}. Vectorial structured light has been suggested to improve resilience because of the invariance of the polarisation degree of freedom, but numerous studies in turbulence \cite{cox2020structured,gu2009scintillation,cheng2009propagation,cai2008average,ji2010propagation,wang2008propagation,Cox:16,lochab2018designer} have been inconclusive, with some reporting that the vectorial structure is stable \cite{lochab2018designer,gu2009scintillation,cheng2009propagation}, and others not \cite{cai2008average,ji2010propagation,wang2008propagation,Cox:16,Ndagano2017}.  Careful inspection of the studies that report vectorial robustness in noisy channels reveal that the distances propagated were short and the strength of perturbation low, mimicking a phase-only near-field effect where indeed little change is expected, and hence these are not true tests for robustness or invariance.  Studies that claim enhanced resilience of vector modes over distances comparable to the Rayleigh length \cite{lochab2018designer,gu2009scintillation} have used the variance in the field's intensity as a measure, a quantity that one would expect to be robust due to the fact that each polarisation component behaves independently and so will have a low covariance. This failing of structured light in turbulence has led to numerous correction techniques, including novel encoding/decoding methods \cite{zhu2019compensation}, modal diversity as an effective error-reduction scheme \cite{cox2018modal}, traditional adaptive optics for pre- and post-correction \cite{tyson2002bit,zhao2012aberration,ren2014adaptive} as well as vectorial adaptive tools \cite{he2020vectorial}, iterative routines \cite{li2017gerchberg} and deep learning models~\cite{liu2019deep}.
\begin{figure}[h!]
    \centering
    \includegraphics[width= \linewidth]{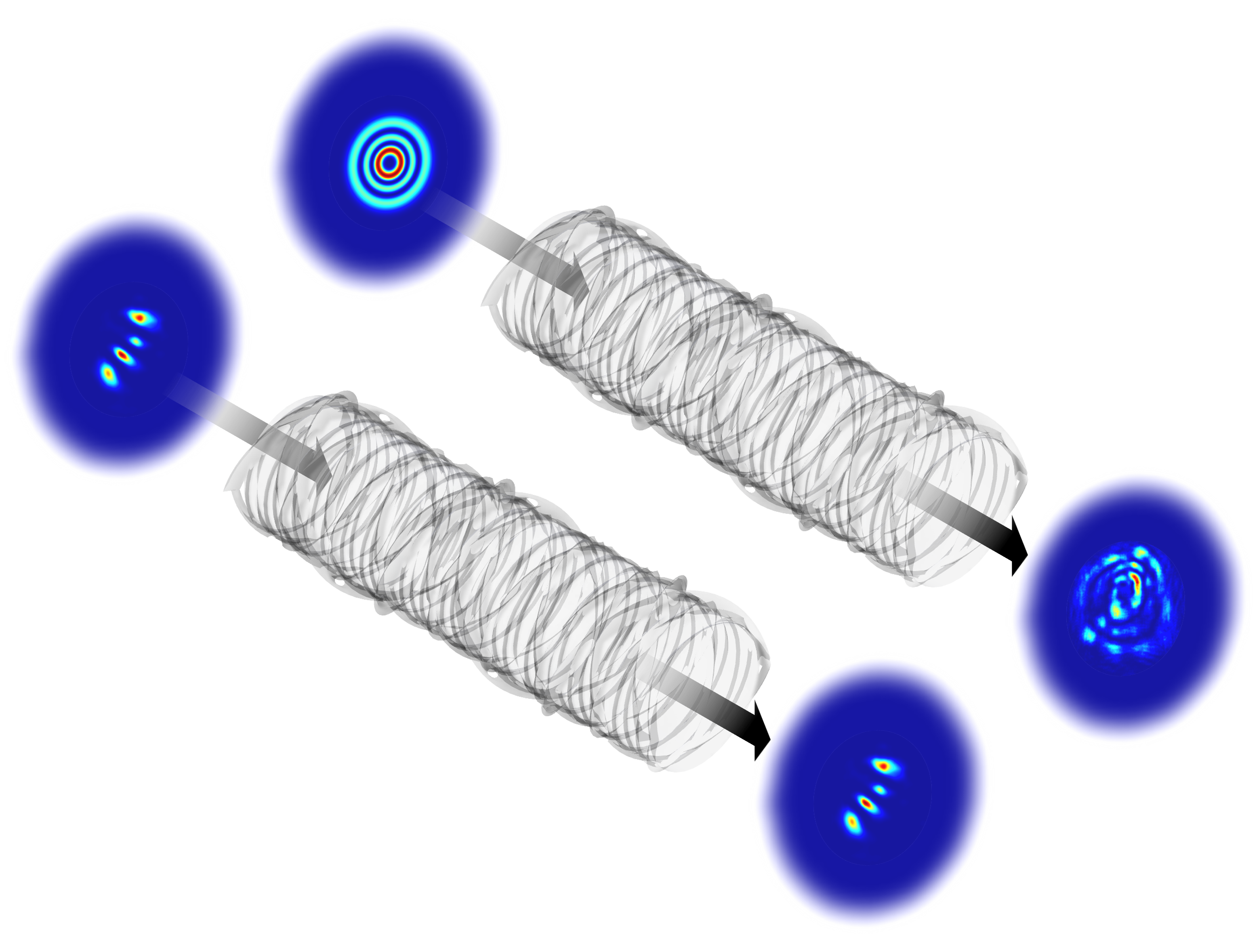}
    \caption{\textbf{Propagation through turbulence.} Most common forms of structured light (such as Laguerre-Gaussian modes) become distorted when propagating through the atmosphere due to the effects of atmospheric turbulence. Compare this to an eigenmode of atmospheric turbulence, which remains unchanged when propagating through the same channel. }
    \label{fig:concept1}
\end{figure}

Here we present a class of structured light whose entire structure in amplitude and phase remains invariant as they propagate through a turbulent free-space channel. We deploy an operator approach to  find the eigenmodes of atmospheric turbulence, a significant departure from prior phenomenological approaches.  Unlike other spatial modes, these exotically structured eigenmodes need no corrective procedures and are naturally devoid of deleterious effects such as modal crosstalk. Moreover, they are valid over any path length in the medium so long as the medium conditions remain static over the time frame of the beam transport, always true for atmospheric turbulence (which typically changes at the Greenwood frequency of 100s of Hz, much slower than the speed of light). We demonstrate this invariance numerically and confirm it experimentally with a laboratory simulated long path comprising weak, medium and strong turbulence, implemented using multiple turbulent phase screens along the propagation path.  Our approach offers a new pathway for exploiting structured light in turbulence, and can be easily extended to arbitrary noisy channels whose characteristics are known. 

\begin{figure*}[t!]
    \centering
    \includegraphics[width= \linewidth]{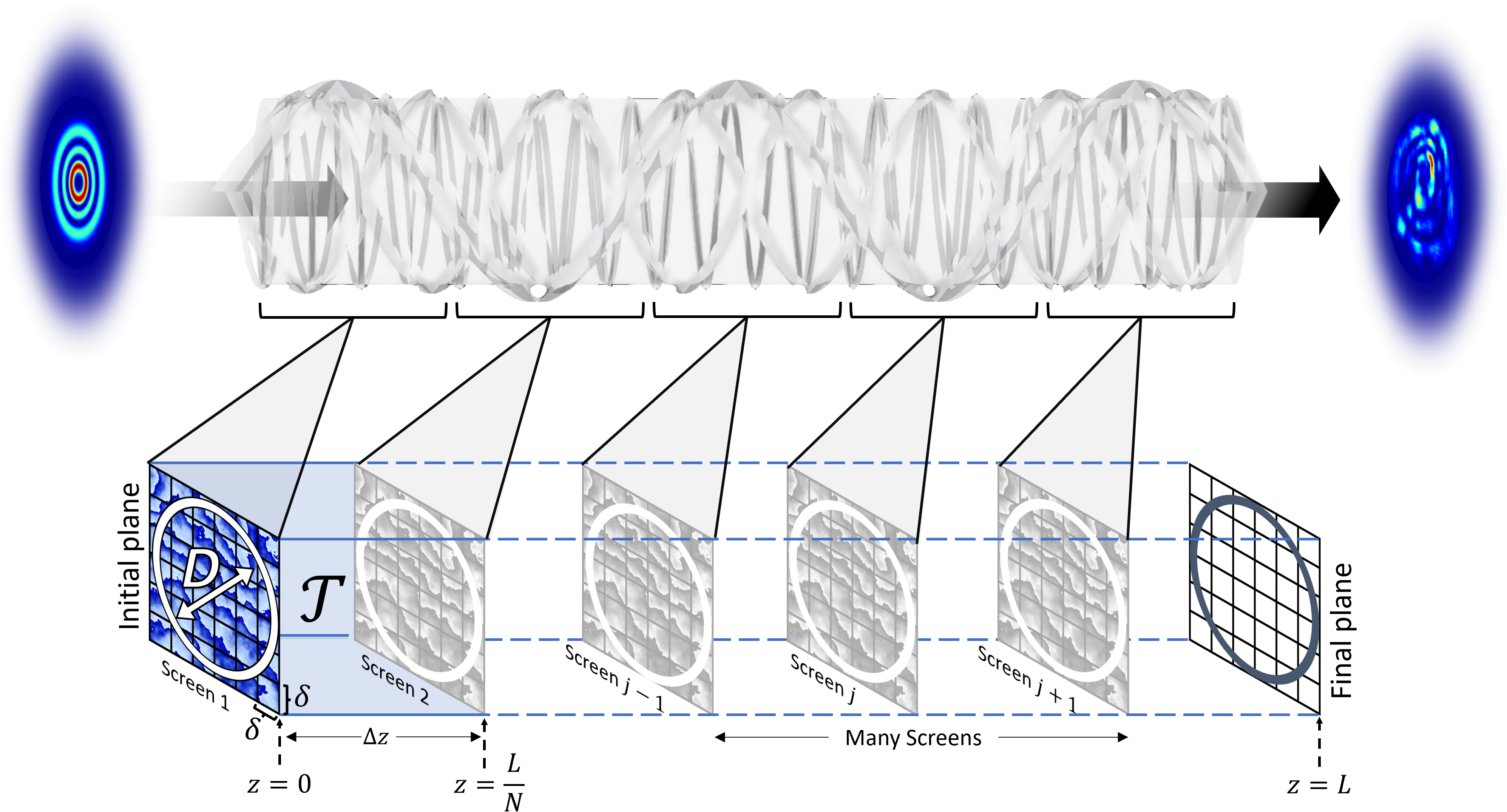}
    \caption{\textbf{The unit cell.}
    The first turbulent screen is placed at the beginning of the channel, at $z = 0$, with subsequent screens placed a distance $ \Delta z = L/N $ away from the prior, where $N$ is the number of turbulent phase screens used.  Each phase screen and distance form a unit cell, the first highlighted in blue, forming $N$ unit cells over the complete path length of $z = L$. The operator for each unit cell, $\mathcal{T}$, is identical, so we need only consider the first unit cell.  The initial plane is discretised into pixels with side length $\delta$ and turbulence is simulated with a strength characterised by the ratio $D/r_0$, where $D$ is the aperture of the inscribed circle and $r_0$ is Fried's parameter.  The operator describes the action of an imprinted turbulent phase on the beam, followed by vacuum propagation over a distance $\Delta z$.}
    \label{fig:concept2}
\end{figure*}

\section{The eigenmodes of turbulence}
The concept we tackle here is illustrated in Fig.~\ref{fig:concept1}.  Structured light is typically distorted after propagation through free-space due to atmospheric turbulence.  In contrast, the eigenmodes of turbulence are complex forms of structured light that are invariant to the channel, emerging distortion free. To solve for these eigenmodes, we use the multiple screen approximation to Fig.~\ref{fig:concept1} as shown in Fig.~\ref{fig:concept2}.  Here we outline the theory and its implications, before moving on to numerical and experimental confirmation.

\subsection{Theory}
\label{sec:theory}
\begin{figure*}[t!]
    \centering
    \includegraphics[width=0.9 \linewidth]{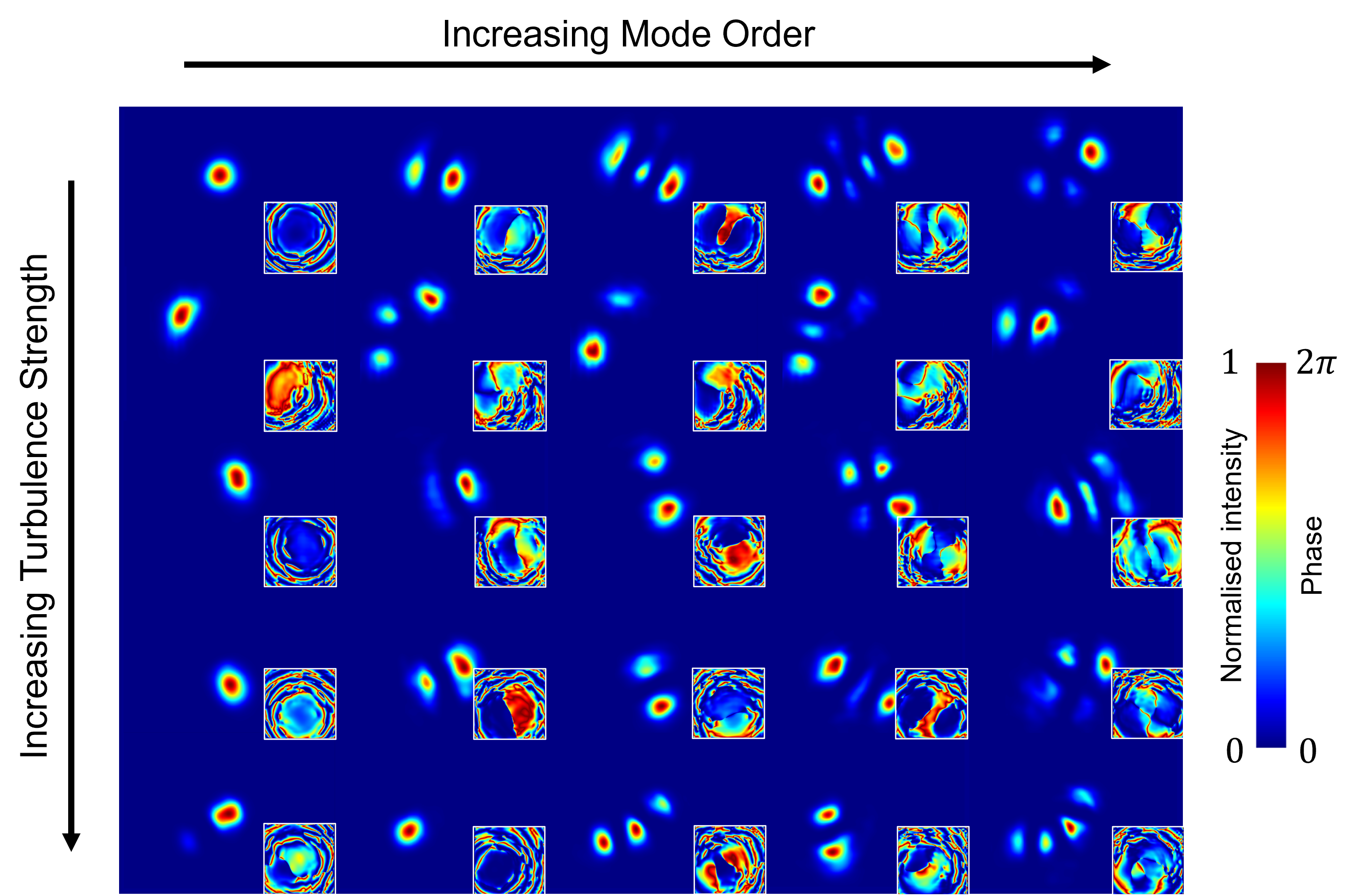}
    \caption{\textbf{Eigenmodes of turbulence.} The numerically calculated eigenmodes of turbulence, showing the first five modes (columns) as a function of turbulence strength (rows).  The insets show the phase profile.  All eigenmode were calculated for a total propagation path of 100 m through weak, medium and strong turbulence as defined by the Rytov variance ($\sigma^2_R$) and Fried parameter ($r_0$). The first two rows show eigenmodes of weak turbulence with $\sigma^2_R = 0.5$ and $r_0 = 1.8 \text{ mm}$. The next two rows show eigenmodes of medium turbulence with $\sigma^2_R = 1$ and $r_0 = 1.2 \text{ mm}$. The last row shows eigenmodes of strong turbulence with $\sigma^2_R = 1.5$ and $r_0 = 0.91 \text{ mm}$.}
    
    \label{fig:numerical-modes}
\end{figure*}
The effects of turbulence are mathematically captured in the stochastic refractive index $n = 1+\delta n$, where $\delta n$ is the random variation in the refractive index of the Earth's atmosphere. It is assumed that $\delta n$ has a zero mean value, i.e. $\left< \delta n \right> = 0$, and that the variation is small, so $|\delta n| \ll 1$. The introduction of this varying term produces the stochastic paraxial Helmholtz equation for a field $\text{V}(x,y,z)$ 
\begin{equation}\label{eq:paraxial helmholtz equation}
    \left(\nabla^2_t + 2ik\partial_z + 2k^2 \delta n \right)\text{V} = 0,
\end{equation}
where $\nabla^2_t$ is the transverse Laplacian and $k = 2\pi/\lambda$ is the wavenumber for wavelength $\lambda $. Equation \ref{eq:paraxial helmholtz equation} can be solved numerically according to the split-step method~\cite{schmidt2010numerical}, illustrated in Fig.~\ref{fig:concept2}.  Multiple random phase screens are placed at equal distances along the beam's propagation path. Importantly, each screen is in the weak turbulence limit and contributes a random phase $\Theta_j$, where $j$ labels the $j$th screen, so that a single screen approximation is valid, but the sum of many such screens can lead to medium or even strong turbulence. 

We introduce an operator approach to the problem. We exploit the fact that the path is subdivided into identical units, each containing such a single screen and a zero turbulence propagation path of length $\Delta z$.  We realise that in the language of operators, the action of each unit $\hat{\mathcal{U}}$ on some field V is given by the product of the operators $\hat{\mathcal{P}}\hat{\mathcal{T}}$, where $\hat{\mathcal{P}}$ and $\hat{\mathcal{T}}$ refer to free-space propagation and turbulence respectively. The action of $\hat{\mathcal{U}}$ on a field V is given by the Huygen-Fresnel integral with a turbulent phase factor
\begin{equation}\label{eq:huygen-fresnel integral with phase factor}
    \hat{\mathcal{U}} \text{V} \equiv \int\dd[2]{\mathbf{x}}\; g(\mathbf{x},\mathbf{x}'; \Delta z)\exp(i\Theta) \text{V}(\mathbf{x},z=0) ,
\end{equation}
where
\begin{equation}\label{eq:green's function}
    g(\mathbf{x},\mathbf{x}'; \Delta z) = \frac{1}{i\lambda \Delta z} \exp\left(\frac{i\pi}{\lambda  \Delta z}\norm{\mathbf{x}-\mathbf{x}'}^2 \right)
\end{equation}
is the paraxial free space Green's function and $\mathbf{x}$, $\mathbf{x}'$ are the two-dimensional coordinates of the initial and final planes, respectively. We then discretize $\mathbf{x}$ and $\mathbf{x}'$ into grids of $N \times N$ points. The coordinates are labelled $\mathbf{x} \equiv (x_\alpha,y_\beta)$ and $\mathbf{x}' \equiv (x_\mu,y_\nu)$, so that $\hat{\mathcal{U}}$ is given by 
\begin{equation}\label{eq:explicit form of operator U}
\begin{gathered}
    \mathcal{U}_{\mu\nu\alpha\beta} = \frac{1}{i\lambda \Delta z} \exp\left(\frac{i\pi}{\lambda \Delta z }\left(x_\mu-x_\alpha\right)^2\right) \\ 
    \times\exp\left(\frac{i\pi}{\lambda \Delta z }\left(y_\nu-y_\beta\right)^2\right) \exp\left(i\Theta(x_\alpha,y_\beta)\right).
\end{gathered}
\end{equation}
An eigenmode E is then a solution to the tensor eigenvalue equation
\begin{equation}\label{eq:eigenvalue equation}
    \gamma_n \text{E}^n_{\mu\nu}  = \mathcal{U}_{\mu\nu\alpha\beta} \text{E}^n_{\alpha\beta},
\end{equation}
where $\gamma_n$ is the eigenvalue of the nth eigenmode. Repeated indices are implicitly summed over and $\text{E}_{\mu\nu} \equiv \text{E}(x_\mu,y_\nu)$. To convert the above tensor equation into the usual matrix-vector form, we specify a mapping $\rho$ that acts on the indices $(\alpha,\beta)$ and $(\mu,\nu)$ and ``counts'' them, first by columns and then by rows, such that $\rho(1,1) = 1$, ..., $\rho(N,1) = N$, $\rho(1,2) = N+1$ up to $\rho(N,N) = N^2$. This mapping lets us rewrite Eq.~\ref{eq:eigenvalue equation} as 
\begin{equation}\label{eq:matrix-vector form of eigenvalue equation}
    \gamma_n \text{E}^{n}_i = \mathcal{U}_{ij} \text{E}^n_{j},
\end{equation}
since $\rho(\alpha,\beta) = j$ and $\rho(\mu,\nu ) = i$. This equation can be routinely solved using numeric methods to find the eigenmodes of the unit cell operator. The action of the full channel is then described by the product $\hat{\mathcal{U}}_n...\hat{\mathcal{U}}_1$ of repeated unit cells, and as per the definition of eigenmodes, they remain invariant regardless of the number of operators applied.  

Turbulence is a stochastic process in which the refractive index of the Earth's atmosphere varies according to well-known statistics, having zero mean and some non-zero variance.  To see the impact of averaging over many different instances of turbulence on the robustness of modes, we return to the the Helmholtz equation but this time in the more general non-paxarial form
\begin{equation}\label{eq:non-paraxial helmholtz equation}
    \left(\nabla^2 + k^2\right)\text{V} = -2k^2\delta n \text{V}, 
\end{equation}
which has the solution
\begin{equation}\label{eq:solution to non-paraxial helmholtz equation}
    \text{V}(\mathbf{r}') = 2k^2\int\dd[3]{\mathbf{r}}\; G(\mathbf{r},\mathbf{r}')\text{V}(\mathbf{r})\delta n(\mathbf{r}),
\end{equation}
with $G(\mathbf{r},\mathbf{r}') = \exp(ik\norm{\mathbf{r}-\mathbf{r}'})/4\pi\norm{\mathbf{r}-\mathbf{r}'}$ being the free-space Green's function and $\mathbf{r} = (\mathbf{x},z)$. Taking the ensemble average and using the result that $\left<\delta n \text{V}\right> = \mathcal{A} \left< \text{V} \right>$~\cite{andrews2005laser}, we find
\begin{equation}\label{eq:average solution to non paraxial helmholtz equation}
    \left<\text{V}(\mathbf{r}')\right> = 2k^2 \mathcal{A} \int\dd[3]{\mathbf{r}}\; G(\mathbf{r},\mathbf{r}')\left<\text{V}(\mathbf{r})\right>,
\end{equation}
\noindent where the constant $\mathcal{A}$ is related to the covariance of the refractive index fluctuations. 

%There are two ways to interpret Eq.~\ref{eq:average solution to non paraxial helmholtz equation}. Firstly, we recognize that Eq.~\ref{eq:average solution to non paraxial helmholtz equation} is identical to the usual, zero-turbulence Fresnel integral, up to a constant.  Therefore, the averaged eigenmodes should be solutions to the free-space, no turbulence, case.  In other words, if the channel involves some form of averaging, say at the detector, then the best mode set in this case is identically the traditional free-space modes in various geometries: Hermite-Gaussian, Laguerre-Gaussian and so on.  Secondly, if we substitute a free-space mode in the initial plane ($z = 0$) instead of $\left< \text{V}(\mathbf{r}) \right>$ as the driving term, we can conclude that there is no average effect/distortion of turbulence on free-space modes.  
We recognize that Eq.~\ref{eq:average solution to non paraxial helmholtz equation} is identical to the usual, zero-turbulence Fresnel integral, up to a constant.  Therefore, the averaged eigenmodes should be solutions to the free-space, no turbulence, case.  In other words, if the channel involves some form of averaging, say at the detector, then the best mode set in this case is identically the traditional free-space modes in various geometries: Hermite-Gaussian, Laguerre-Gaussian and so on.  

\subsection{Numerical simulation}
\begin{figure}
  \centering
   \includegraphics[width = \linewidth]{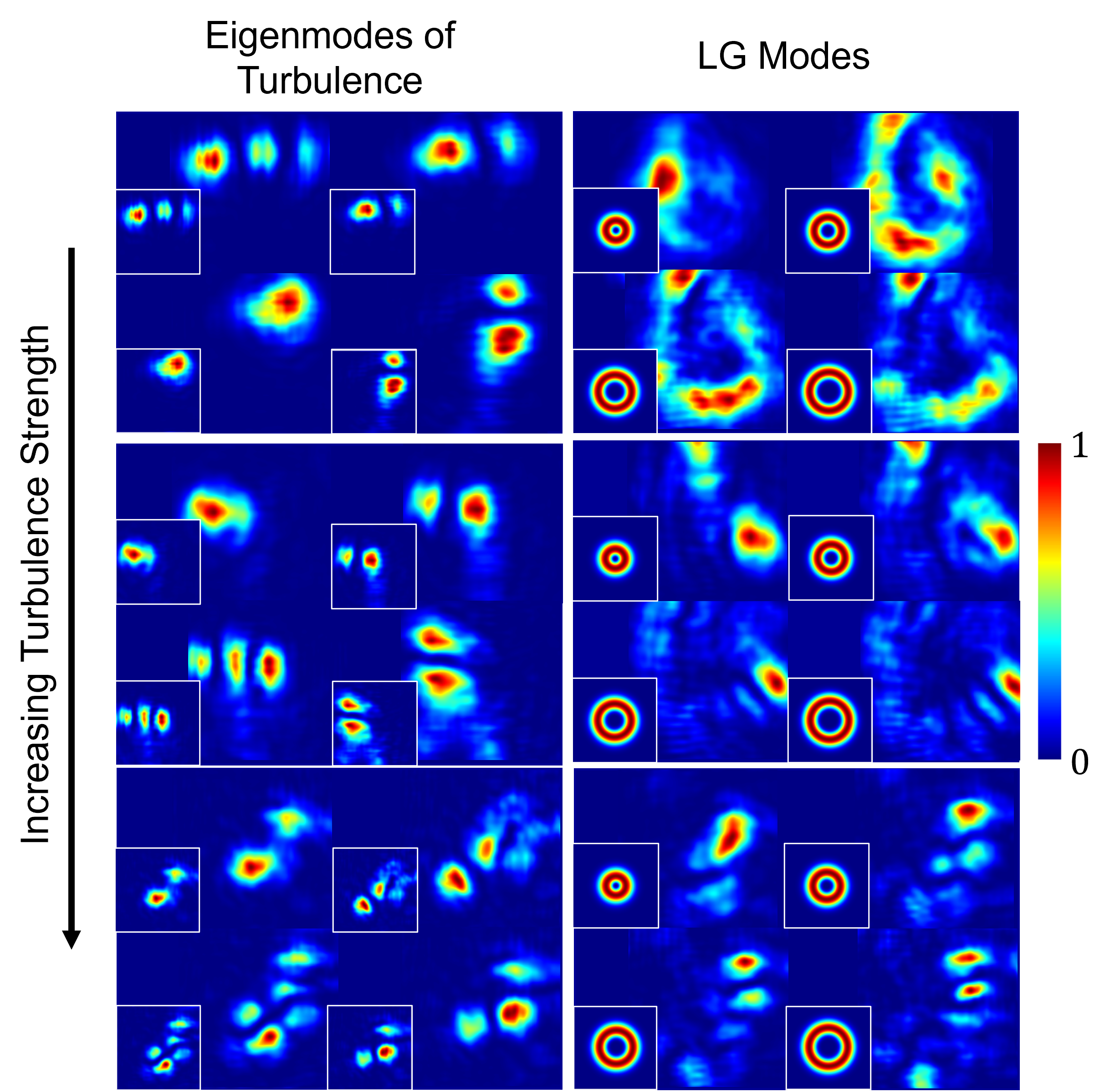}
    \caption{\textbf{Invariance of the eigenmodes.} The eigenmodes (left) and LG modes (right) after propagation through weak, medium and strong turbulence  through a channel equivalent to propagating over a distance of 100 m. The insets shows the modes before experiencing turbulence. The numerical simulations used the split step method with three unit cells each consisting of a turbulence screen with a given $r_0$ followed by $33.33 \text{ m}$ of propagation. Weak turbulence was characterised by $\sigma^2_R = 0.5$ and $r_0 = 1.8 \text{ mm}$, medium turbulence by $\sigma^2_R = 1$ and $r_0 = 1.2 \text{ mm}$  and strong turbulence by $\sigma^2_R = 1.5$ and $r_0 = 0.91 \text{ mm}$}
   \label{fig:num-prop}
\end{figure}

\begin{figure}[h!]
    \centering
    \includegraphics[width=0.9\linewidth]{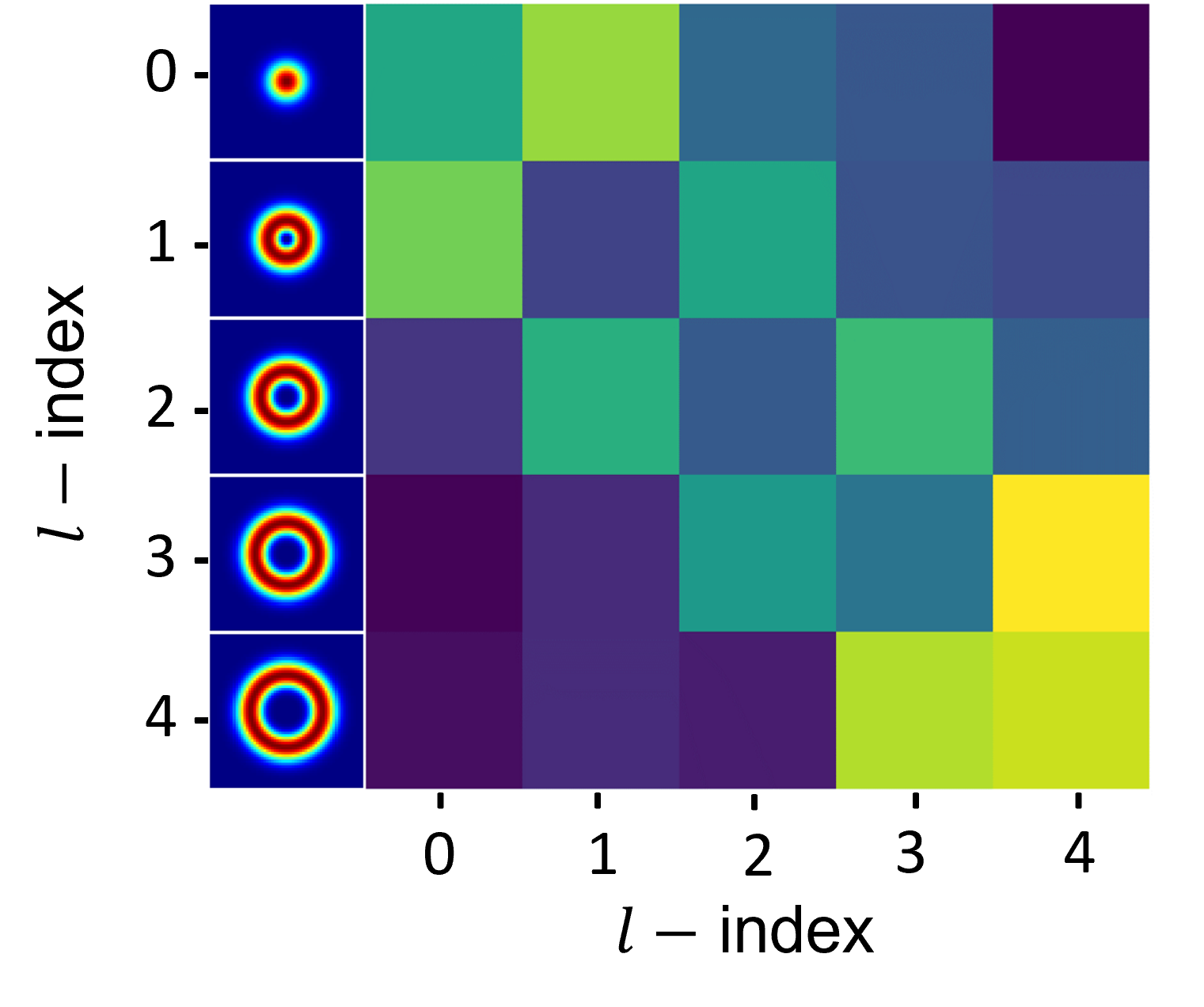}
    \includegraphics[width=0.9\linewidth]{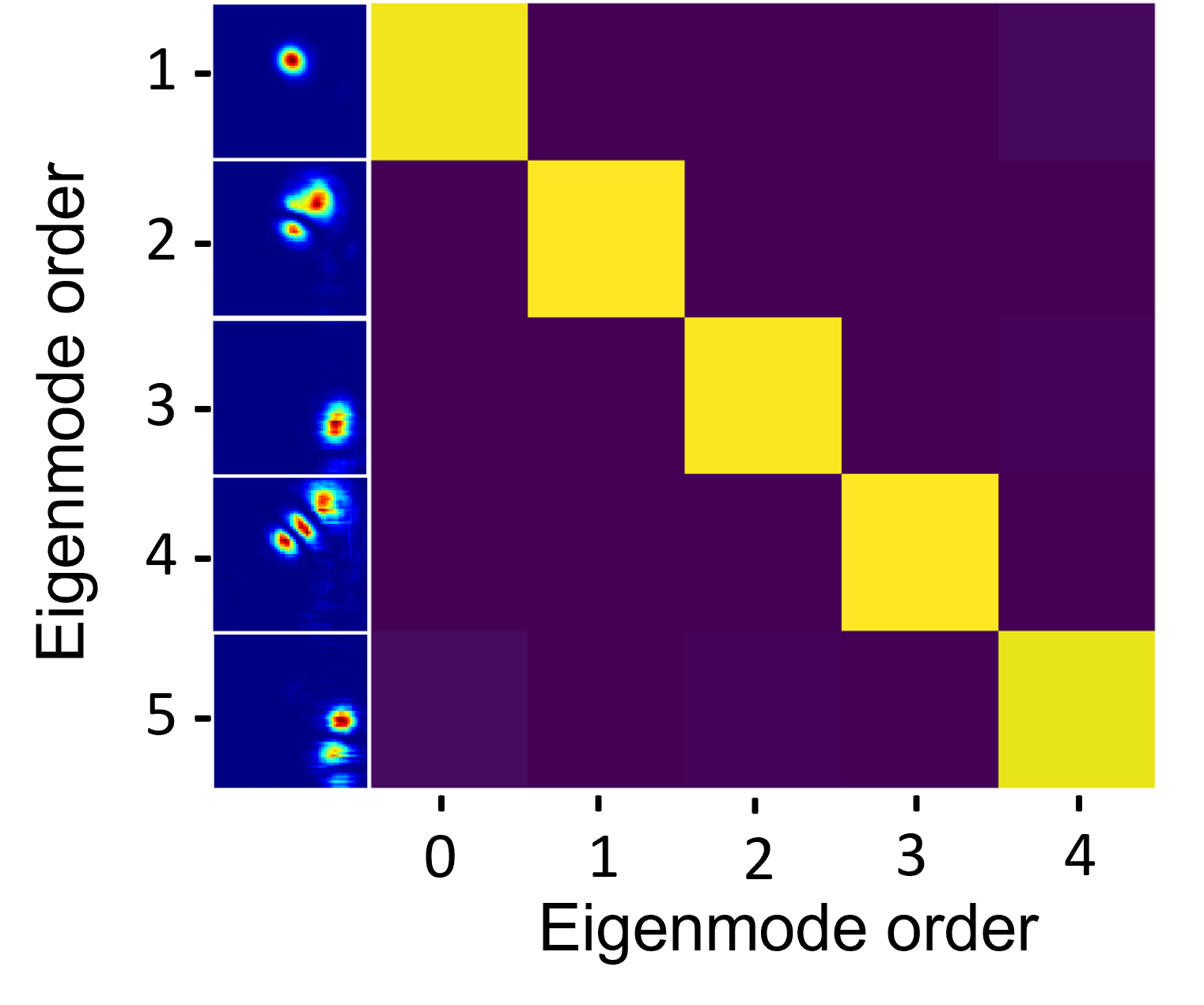}
    \caption{\textbf{Cross talk free transmission.} Cross talk matrices for OAM modes $\ell \in [0,4]$ (top) and eigenmodes (bottom) with insets showing the intensity of the beams. The eigenmodes are unchanged and remain orthogonal, whereas the OAM modes scatter into each other. Turbulence results shown for $D/r_0 = 2$ with a total path length of 100 m and a beam waist parameter for the OAM beams of $w_0 = 6.67 \text{ mm}$.}
    \label{fig:crosstalk}
\end{figure}

To validate the theory, we first use the operator to calculate the eigenmodes, and then numerically propagate them through a thick medium of atmospheric turbulence using the split-step approach illustrated in Fig.~\ref{fig:concept2}.  Throughout the text, for clarity and brevity, we show only the low order eigenmodes and use the OAM modes as our point of comparison.   Examples of the intensity and phase of the eigenmodes are shown in Fig.~\ref{fig:numerical-modes}.  Here the first five eigenmode solutions are shown in the left column, increasing from left to right, with the rows corresponding to the turbulence strength, increasing from top to bottom.  Although these are complex forms of structured light, as eigenmodes to turbulence they should be invariant after propagation through a turbulent atmosphere.  To test this, we propagate OAM carrying Laguerre-Gaussian (LG) modes and the eigenmodes through various scenarios of turbulence over a 100 m path length, with the results shown in Fig.~\ref{fig:num-prop}.  We see that while the OAM modes are distorted, the eigenmodes are robust.  This can be quantified by performing a modal analysis \cite{pinnell2020modal} at the end of the turbulent channel, as would be the case in optical communication at the receiver.  In Fig.~\ref{fig:crosstalk} we see that while the crosstalk is substantial for OAM modes when propagated through turbulence, evident from the many off-diagonal terms, the eigenmode crosstalk matrix remains diagonal after the same channel, for minimal crosstalk.

\begin{figure}[h!]
    \centering
    \includegraphics[width=\linewidth]{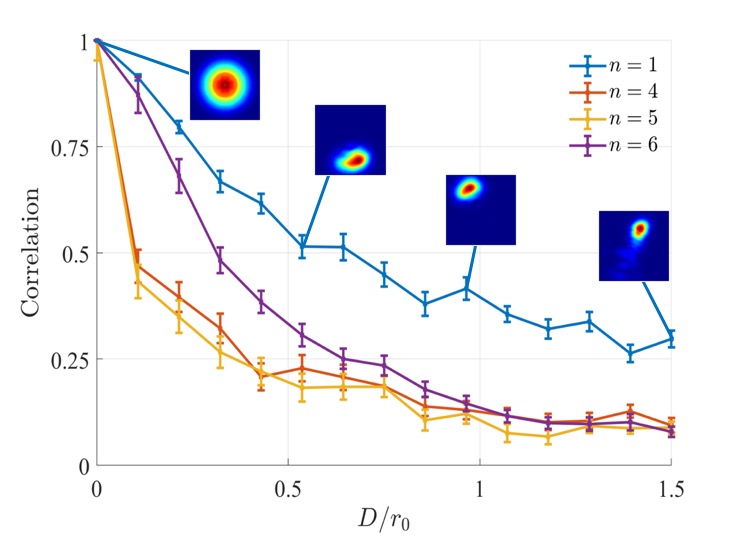}
    \caption{\textbf{Evolving robust structure.} The correlation (overlap integral) for four orders (1, 4, 5 and 6) of eigenmodes with their corresponding vacuum counterparts is shown for a range of turbulence strengths, characterised by $D/r_0$. Insets show intensity patterns for the first order ($n = 1$) eigenmode at various turbulence strengths.}
    \label{fig:compare}
\end{figure}

It is instructive to consider the evolution from the free-space modes to the eigenmode structure as turbulence is steadily increased, shown graphically in Fig.~\ref{fig:compare}. When there is no turbulence ($D/r_0 = 0$) the operator correctly returns the free-space modes, with the inset showing the first order solution - the Gaussian mode.  As turbulence increases so the eigenmode structure changes in amplitude and phase, with insets showing the new first order eigenmodes.  The correlation of the new eigenmodes to the initial free-space modes are plotted versus turbulence strength.  Initially the operator returns the true free-space modes (correlation of 1), but as turbulence increases so the deviation of the eigenmodes from the free-space modes increases, resulting in a decreased correlation.  For higher-order modes (shown up to order 6), the deviation is so great as to make the correlation close to zero: there is little similarity remaining between the free-space mode and the eigenmode of turbulence.  This likely explains why no free-space modal family has been found to be robust to turbulence - they are just ``too far'' in a modal sense from what is required.
\begin{figure}[h!]
    \centering
    \includegraphics[width=\linewidth]{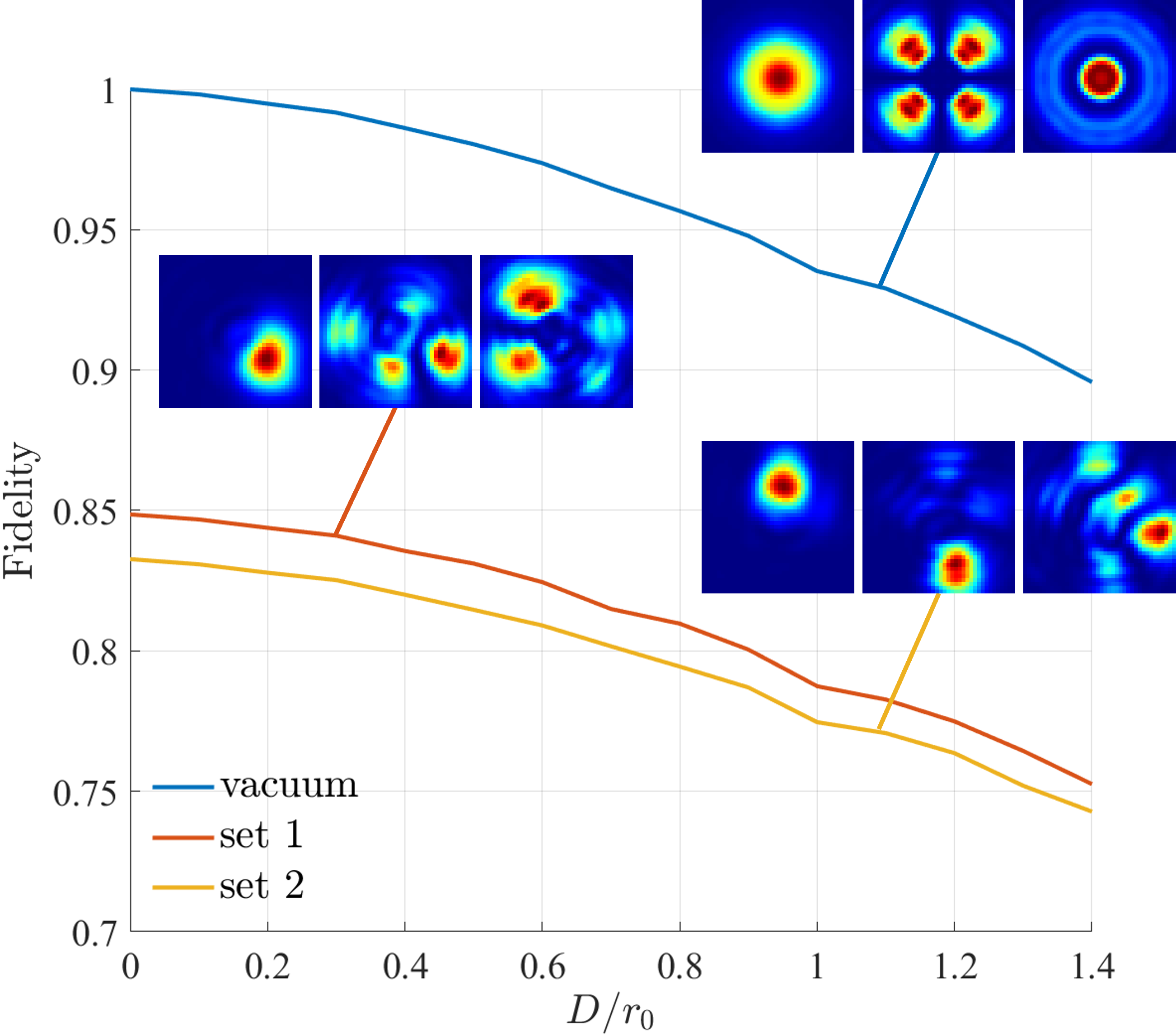}
    \caption{\textbf{Modal averaging.} The averaged fidelity as a function of turbulence strength for vacuum and two examples of eigenmodes (set 1 and set 2). The eigenmodes are calculated for a particular instance of turbulence but used in conditions that differ from this. The fidelity is calculated as the average value of the elements along the diagonal of their $4\times 4$ cross talk matrices. While the vacuum modes are perfectly orthogonal in the absence of turbulence (fidelity of 1), their fidelity decreases for increasing turbulence strength. The turbulent eigenmodes are not orthogonal in the absence of turbulence, and their fidelity is consistently worse than the vacuum modes under this averaging effect.}
    \label{fig:average}
\end{figure}

In contrast, if the conditions are varying faster than the eigenmodes can be altered, and the receiver averages the output from the channel, then as predicted by our theory, the free-space modes outperform the eigenmodes, as shown in Fig.~\ref{fig:average}.  Here, the robustness of the performance of the eigenmodes is examined when the turbulence conditions are changed and the detection averaged.  The many different instances of turbulence exclude the phase screens used to form the modes initially. Four beams (mode orders $n = 1, 4, 5, 6$) were sent through a turbulent channel and their cross talk quantified using the average value of the elements along the diagonal of their $4\times 4$ crosstalk matrices. This average, plotted as the fidelity, shows that while the vacuum modes are perfectly orthogonal in the absence of turbulence, their fidelity decreases (greater scattering into other modes) for increasing turbulence strength. However, the eigenmodes of turbulence are not orthogonal in the absence of turbulence, and their fidelity is consistently worse than the vacuum modes when passed through a medium for which they were not designed.  This highlights an important aspect of the eigenmodes - they are robust so long as the medium for which they were created is valid.

\section{Experimental Results}
\begin{figure*}[t]
    \centering
    \includegraphics[width=\linewidth]{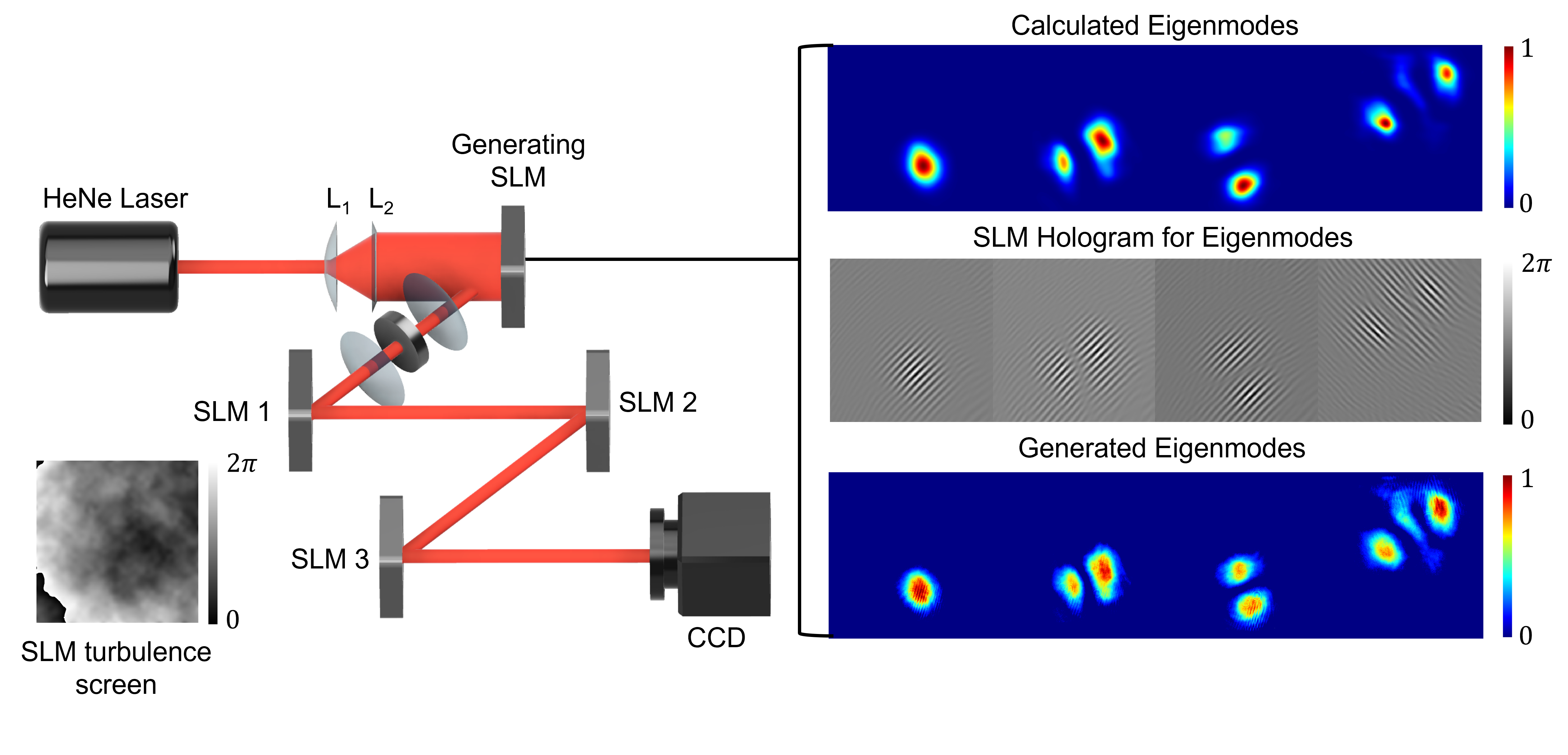}
    \caption{\textbf{Experimental setup.} Lenses L$_1$ and L$_2$ expand and collimate a laser beam onto an SLM on which a hologram of the initial beam is displayed. The ideal, turbulence-free beam is generated at this plane and subsequently propagates through three turbulent screens which are also displayed on SLMs, each followed by 1 m of free-space propagation. The final aberrated field is captured on a CCD to image its intensity. The panels on the right show examples of the desired eigenmodes (Calculated Eigenmodes), the holograms to create them, and the measured eigenmodes without any turbulence or propagation (Generated Eigenmodes).}
    \label{fig:exp}
\end{figure*}
The experiment, shown in Fig.~\ref{fig:exp}, is conceptually divided into three parts. In the generation stage, a He-Ne laser beam (wavelength $\lambda = 633$ nm) was expanded using a 10$\times$ objective lens L$_1$ and then collimated by L$_2$ ($f_2$ = 150 mm) before being directed onto a reflective PLUTO-VIS HoloEye spatial light modulator (SLM) which generated the desired initial field. This field then entered the turbulent section of the setup where it passed through three unit cells, each comprising the same random phase screen and a propagation distance of one metre. The phase screens were generated using the sub-harmonic random matrix transform method~\cite{schmidt2010numerical} and displayed on the SLMs. This perturbed field was then detected and measured on a camera (CCD). The panels on the right of the experimental setup in Fig.~\ref{fig:exp} show four examples of the desired (calculated) eigenmodes, the holograms to create them by complex amplitude modulation, and the experimental validation that without any turbulence or propagation, that they are created (generated eigenmodes) with high fidelity (bottom panel).

%In the eigenmode generation stage, a He-Ne laser (wavelength $\lambda = 633$ nm) was expanded using a 10$\times$ objective lens L$_1$ and then collimated by L$_2$ ($f_2$ = 150 mm) before being directed onto a reflective PLUTO-VIS HoloEye SLM (8 \textmu m, 1920$\times$1080 pixels, calibrated for a 2$\pi$ phase shift for $\sim$633 nm).  %The created beam was then passed through three unit cells of programmed turbulence and a distance of 1 m, using successive SLMs, before being imaged by a CCD for measurement. 

Our setup differs from conventional laboratory simulations of turbulence in that we are able to mimic a thick path, from weak to strong turbulence, whereas often only a single phase screen is used, allowing only weak turbulence to be tested.  Using our setup, we studied an effective real-world channel of $L = 100 \text{ m}$, at our wavlength of $\lambda  = 633 \text{ nm}$ and with Rytov variances of $\sigma^2_R = 1.5,\, 1 \text{ and } 0.5$, corresponding (respectively) to strong, medium and weak turbulence, with Fried parameters ($r_0$) of $0.47\text{ mm, } 0.62\text{ mm and } 0.93\text{ mm}$, respectively. We required three screens for each turbulence strength, separated by a distance of $33.3$ m, each with effective  Fried parameters $r_{0,s} = 0.9\text{ mm, } 1.2\text{ mm and } 1.8\text{ mm}$ while maintaining a Rytov variance in each slab (segment of the channel) to be smaller than 0.9, 0.6 and 0.3, respectively. This channel was simulated on the setup shown in Fig.~\ref{fig:exp} using the Fresnel scaling procedure \cite{rodenburg2014simulatingnew}, allowing a long path to be generated within laboratory dustances. The scaling factors were chosen to be: $\alpha_x = \alpha_{x'} = \sqrt{0.03} \approx 0.173 $ and $\alpha_z = 0.03$. This corresponded to a total path length of $L'= 3\text{ m}$ and segment Fried parameters of $r_{0,s} = 0.081\text{ mm, } 0.11\text{ mm and } 0.16\text{ mm}$ (see Appendix for details).

%To test this experimentally, we built the setup shown in Fig.~\ref{fig:exp}, which comprises two parts, an eigenmode generation stage and a turbulence stage.  In the former, complex amplitude modulation on a spatial light modulator (SLM) is used to create the desired modes, while in the latter, three unit cells are programmed on SLMs for an effective turbulence that can vary from weak to strong, covering an effective channel of 200 m in a medium with a Rytov variance of approximately 1.  

\begin{figure}
    \centering
    \includegraphics[width=\linewidth]{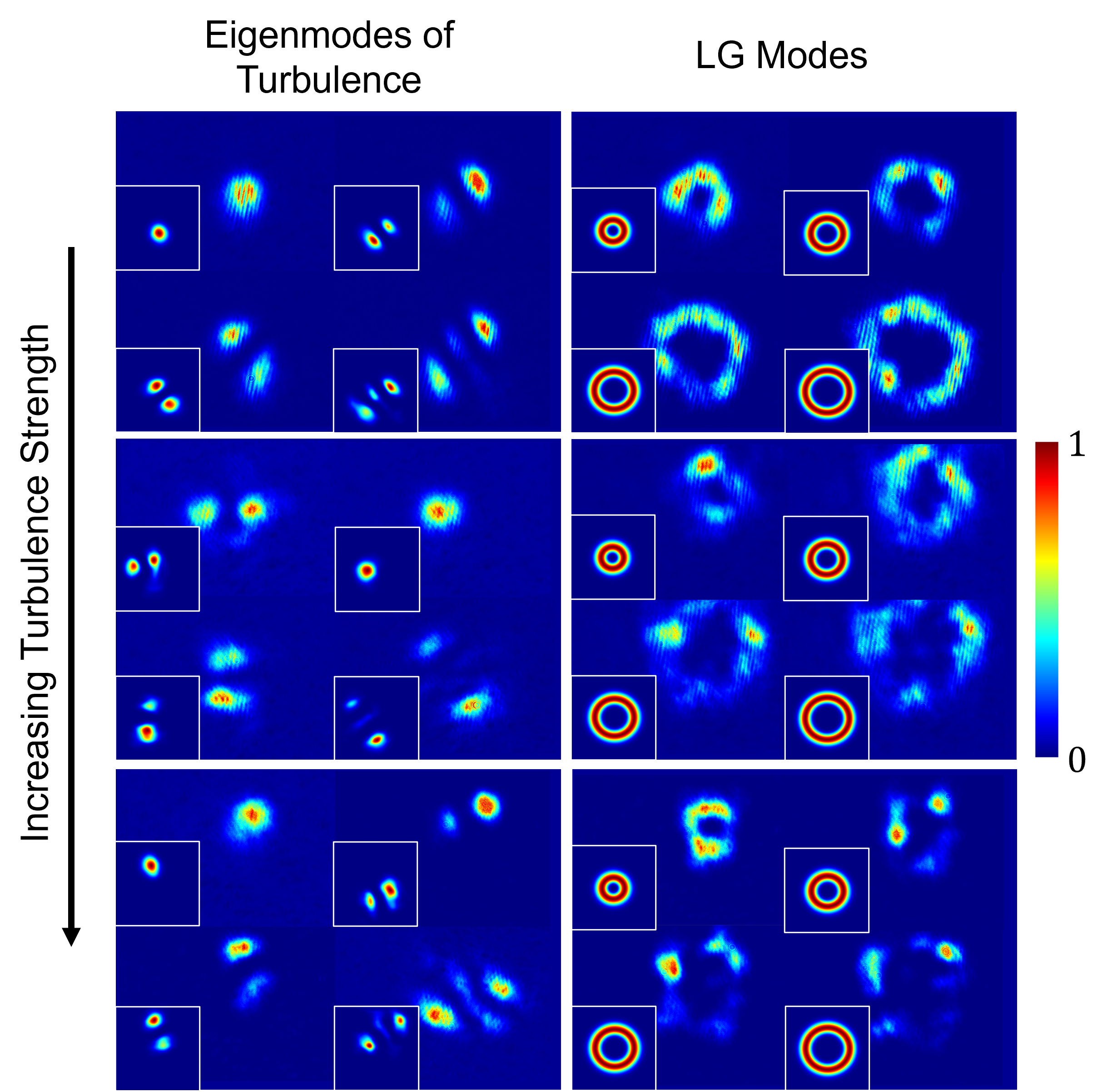}
    \caption{\textbf{Experimental eigenmodes.} In the left column we see the measured intensities of the eigenmodes of turbulence after propagating through the experimental setup.  The results show eigenmodes of weak, medium and strong turbulence.  In the right column we see the measured intensities of OAM modes after propagating through the same experimental setup with the same turbulence phase screens for comparison. The insets show the initial input mode. Weak turbulence was characterised by $\sigma^2_R = 0.5$ and $r_0 = 1.8 \text{ mm}$, medium turbulence by $\sigma^2_R = 1$ and $r_0 = 1.2 \text{ mm}$  and strong turbulence by $\sigma^2_R = 1.5$ and $r_0 = 0.91 \text{ mm}$ }
    \label{fig:mainresult}
\end{figure}

The results of OAM and the eigenmodes for weak, medium and strong turbulence are shown in Fig.~\ref{fig:mainresult}.  The collage shows the final measured eignemodes after the channel, with the insets showing the initial mode as prepared prior to the channel.  The robustness of the eigenmodes is clearly evident, in contrast to the highly distorted OAM modes.

\section{Discussion}
%We have developed and implemented a procedure for finding modes which are unchanged after propagating through a turbulent channel by discretising the Fresnel integral into an operator which can be represented in a discrete matrix. These modes are eigenmodes in their truest sense, i.e,. they are fixed under the action of the channel.  This differs from a singular value decomposition procedure to obtain so-called singular modes. Such modes have singular values of one, which means that all the light present in the transmitting plane is collected and detected in the detecting plane, and has been executed with a discretised Green's function \cite{miller2017better} and with two different basis sets \cite{shatokhin2020spatial}, where the input plane is described by LG modes and the output plane is represented in the pixel basis. Such a treatment allows the description of a larger grid size but does not return eigenmodes: even though these modes have unit singular values and thus preserve the light, their field patterns change between the transmitting and receiving planes. 
We have developed and implemented a procedure for finding modes which are unchanged after propagating through a turbulent channel by discretising the Fresnel integral into an operator which can be represented in a discrete matrix. These modes are eigenmodes in their truest sense, i.e,. they are fixed under the action of the channel.  This differs from a singular value decomposition procedure \cite{shatokhin2020spatial} which does not return eigenmodes, but instead requires two different basis sets, one for the input plane and another for the output plane, with the result that the field patterns change between the transmitting and receiving planes. 
%Pai \emph{et al}. developed `scattering invariant modes' by studying a layer of zinc oxide nanopowder which was deposited on a glass slide~\cite{pai2021scattering}.  These modes are solutions to the generalised linear eigenvalue problem $ \mathcal{T}\text{E} = \gamma_\text{air}\mathcal{T}_\text{air}\text{E} $ where $\mathcal{T}$ and $\mathcal{T}_\text{air}$ are the scattering and free-space channels, respectively. The scattering invariant modes $\text{E}$ are not eigenmodes, however, they simultaneously diagonalise the scattering and vacuum operators, and so these modes appear the same after propagating through the scattering medium or free space. This is a weaker condition since the modes are allowed to evolve over propagation, unlike the modes presented here. 

A natural feature of the eigenmodes is that they are channel specific. To be useful in a real-world setting, the transmission should be faster than the time frame over which the turbulence changes and the slow time evolving turbulence (typically 100s of Hz) would have to be monitored to determine the time evolution of the eigenmodes for a continuous transmission channel. One can surmise that machine learning would be ideally suited to such a task. 

\section{Conclusion}
The search for states of structured light that are robust to atmospheric turbulence is a pressing challenge, promising enhanced channel capacity and reach in free-space optical links. Here, we have outlined a theoretical approach to finding the complex forms of structured light which are invariant under propagation through the atmosphere, the true eigenmodes of turbulence, and confirmed its validity both numerically and experimentally. These exotically structured eigenmodes need no corrective procedures, are naturally devoid of deleterious effects and are valid over any path length in the medium so long as the medium conditions remain valid. Our approach offers a new pathway for exploiting structured light in turbulence, and can be easily extended to other noisy channels, such as underwater and optical fibre.

\section*{Disclosures}
The authors declare no conflicts of interest. 

\section*{Acknowledgments}
Andrew Forbes acknowledges funding from the National Research Foundation (NRF) and the CSIR-NRF Rental Pool Programme.

\section*{Code, Data, and Materials Availability}
Code, data and materials are available on request from the corresponding author.

%\bibliography{report}   % bibliography data in report.bib
%\bibliographystyle{spiejour}   % makes bibtex use spiejour.bst

\newpage

\section*{Appendix}

The channel parameters, like path length, are highly restricted in the laboratory setting. This presents an apparent difficulty to experimentally verifying the eigenmodes. However, a scaling procedure exists \cite{rodenburg2014simulating} which allows us to verify real-world channels in the laboratory. This procedure is presented below.

The Fresnel integral for the full (real-world) channel of length $L$ is
\begin{equation}
    \text{U}_f(\mathbf{r},L) = \frac{\exp(ikL)}{i\lambda L}\int\dd[2]\mathbf{r}'\; \text{U}_i(\mathbf{r}')\exp\left(\frac{i\pi}{\lambda L}\norm{\mathbf{r}-\mathbf{r}'}^2\right).
\end{equation}
We then apply the following scaling parameters: $\mathbf{r}_\text{lab} = \alpha_x\mathbf{r},\; \mathbf{r}'_\text{lab} = \alpha_{r'}\mathbf{r}'$ and $L'= \alpha_z L$, where $\mathbf{r}_\text{lab}$ and $\mathbf{r}'_\text{lab}$ are the coordinates used in the experiment. The diffraction integral becomes
\begin{equation}\label{eq:scaled down diffraction integral}
\begin{gathered}
      \text{U}_f\left(\frac{\mathbf{r}_\text{lab}}{\alpha_{r}}\right) = \frac{\exp\left(ik L'/\alpha_z\right)}{i\alpha_{r'}\lambda L'}\int\dd[2]{\mathbf{r}'_\text{lab}} \text{U}_i\left(\frac{\mathbf{r}'_\text{lab}}{\alpha_{r'}}\right)\times\\
      \times\exp\left(\frac{i\pi\alpha_z}{\lambda L'}\norm{\frac{\mathbf{r}_\text{lab}}{\alpha_r}-\frac{\mathbf{r}'_\text{lab}}{\alpha_{r'}}}^2\right).  
\end{gathered}
\end{equation}
To keep the diffraction equivalent with these scaled coordinates we require the Fresnel number 
\begin{equation}\label{eq:Fresnel number}
    F = \frac{\pi D_i D_f}{4\lambda L}
\end{equation}
to be the same in both the full and scaled-down cases, where $D_i$ and $D_f$ are the aperture diameters in the initial and final planes, respectively. This sets $\alpha_r\alpha_{r'} = \alpha_z$ and the diffraction integral becomes 
\begin{equation}\label{eq:fresnel scaled diffraction integral}
\begin{gathered}
        \frac{\exp\left(ikL'\left(1-\frac{1}{\alpha_z}\right)\right)}{\alpha_r} \text{U}_f\left(\frac{\mathbf{r}_\text{lab}}{\alpha_r}\right)= \exp\left(-\frac{i\pi r_\text{lab}^2}{\lambda f_r}\right)  \frac{\exp\left(ikL'\right)}{i\lambda L'}\times \\
        \times
        \int\dd[2]{\mathbf{r}'_\text{lab}}\; \text{U}_i\left(\frac{\mathbf{r'}_\text{lab}}{\alpha_{r'}}\right)\exp\left(-\frac{\pi (r'_\text{lab})^2}{\lambda f_{r'}}\right)\exp\left(\frac{i\pi}{\lambda L'}\norm{\mathbf{r}_\text{lab}-\mathbf{r}'_\text{lab}}\right),
\end{gathered}
\end{equation}
where
\begin{align}\label{eq:effective lens lengths for scaling}
    f_r &= \frac{L'}{1-\alpha_{r'}/\alpha_r}, \\
    f_{r'} &= \frac{L'}{1-\alpha_{r}/\alpha_{r'}}.
\end{align}
Setting $\alpha_r = \alpha_{r'}$ means that the final and initial planes have the same size in the laboratory setting and $f_{r,r'}\rightarrow \infty$ and the final Fresnel integral, ignoring constant phase factors which arise due to scaling, becomes
\begin{equation}
\begin{gathered}\label{eq:final fresnel integral with new scalings}
    \text{U}_f\left(\frac{\mathbf{r}_\text{lab}}{\alpha_r}\right)=\frac{\alpha_r \exp(ikL')}{i\lambda L'} \int\dd[2]{\mathbf{r}'_\text{lab}}\; \text{U}_i\left(\frac{\mathbf{r'}_\text{lab}}{\alpha_{r'}}\right)
    \\
    \times
    \exp\left(\frac{i\pi}{\lambda L'}\norm{\mathbf{r}_\text{lab}-\mathbf{r}'_\text{lab}}\right).
\end{gathered}
\end{equation}
\end{document}